# Carbon Impurities on Graphene Synthesized by Chemical Vapor Deposition on Platinum

*Jinglei Ping*[1,2] *and Michael S. Fuhrer*[1,2]

[1]Center for nanophysics and Advanced Materials, University of Maryland, College Park, MD 20742-4111, USA, [2]School of Physics, Monash University 3800 Victoria, Australia

**We report nanocrystalline carbon impurities coexisting with graphene synthesized via chemical vapor deposition (CVD) on platinum. We observe micron-size island-like impurity layers, which can easily be mistaken for second graphene layers in optical microscopy or secondary electron microscopy (SEM). The island orientation depends on the crystalline orientation of the Pt, as shown by electron backscatter diffraction(EBSD), indicating growth of amorphous carbon below graphene. Dark-field TEM indicates that in addition to uniform single-crystal graphene, our sample is decorated with nanocrystalline carbon impurities with a spatially inhomogeneous distribution. Raman spectra show a large D peak, however electrical characterization shows high mobility (~6,000 cm$^2$/Vs), indicating a limitation for Raman spectroscopy in characterizing the electronic quality of graphene.**

Platinum was recently developed as the catalytic substrate metal for CVD of graphene[1]. While copper, another promising and widely-used catalytic metal, requires low-pressure synthesis conditions[2] to produce large(>100 μm) single-crystal monolayer graphene, Pt substrates could achieve this with ambient-pressure growth conditions. The electrolysis transfer method[1], which is not applicable to Cu due to its high activity, can remove graphene synthesized on Pt free of metal and avoid etchant residues. Particularly, the melting point of Pt, 1768 °C, is 683 °C higher than that of Cu, 1085 °C, providing an opportunity to study carbon deposition at high temperature. There are also intriguing phenomena observed for graphene grown on Pt, such as sizable (~10μm) multi-layer graphene pyramids and wrinkles twisting graphene from Bernal stacking to rhombohedral stacking[3]. However, graphene synthesized on Pt has not been characterized extensively. Here we characterize Pt-synthesized graphene with optical microscopy, scanning electron microscopy (SEM), electron backscatter diffraction (EBSD), transmission electron microscopy (TEM), atomic force microscopy (AFM), Raman spectroscopy, and electrical transport measurements. We identify novel carbon impurity structures that have not been observed in graphene synthesized by other methods. These impurities consist of nanocrystalline $sp^2$ carbon, and can cause considerable confusion in characterizing the layer number and defect concentration by conventional methods.

**Results**

Figure 1a shows an SEM micrograph of the sample on Pt foil, and Figure 1b shows an optical micrograph of the sample after transfer to $SiO_2$/Si. As will be shown below, monolayer graphene covers the whole surface, but optical and SEM micrographs show few-micron-sized islands of increased contrast. Characterization of graphene on the Pt substrate via energy dispersive spectroscopy shows no significant peaks except carbon and Pt (data not shown in this paper). The size, morphology and configuration of the islands, as shown in Figure 1a, depend on the crystalline orientations of the Pt substrate which are distinguished by contrast in SEM images due to the tunneling effect of the secondary electrons. Such correlation will be investigated further via EBSD below in Figure 2.

Typically graphene is transferred onto silicon dioxide substrates facing the same direction as it is on Pt, i.e. the side originally touching the metal surface touches the silicon dioxide surface after transfer. Imaging the topography of graphene transferred in this "normal" way requires removal of the transfer support, typically poly(methyl methacrylate) ( PMMA), in which case PMMA residue may contaminate the image. To investigate the topography of graphene grown on Pt without exposing the surface to be probed to PMMA, we use an "up-side-down" transfer method: we transfer the graphene/PMMA stack to $SiO_2$/Si with the graphene on top, and do not remove the PMMA.

Figure 1c shows an AFM image of graphene/PMMA transferred by the "up-side-down" method. The surface being imaged has had no contact with PMMA. Islands similar in morphology to those seen in Figs. 1a and b can be identified in the AFM image with height of about 0.35 nm to more than 1 nm. We also observed two types of wrinkles in graphene transferred by this method: raised wrinkles which were valleys in graphene on the Pt growth substrate and may represent the Pt metal grain boundaries, and indented wrinkles which were raised when graphene was on Pt, and likely arose from the differential thermal expansion between graphene and Pt[4]. Such narrow wrinkles are not the consequence of the transfer process as graphene adheres tightly to the 300nm thick PMMA and only wrinkles close to that scale can result from transfer.

We identify the orientations of monocrystal Pt via EBSD and image the island layers grown on corresponding Pt grains via SEM. Crystalline orientations can be compeletely described by either the combination of plane direction and crystal direction {h k l}<u v w> or Euler angles ($\phi_1$, $\Phi$, $\phi_2$). Index {h k l} represents atomic arrangements on the cleavage surface, the last Euler angle, $\phi_2$, represents rotation of the crystal in the plane of cleavage surface with respect to the reference direction (RD), provided that Bunge's (passive) description is used. As an example given in Figure 2, maps of inverse pole figure (IPF), Figure 2b, and Euler angles, Figure 2c, are shown for three monocrystals A, B, and C as in Figure 2a. Given {h k l} and $\phi_2$, we obtain the top-view of the crystals shown in Figure 2g-i. We define a direction in the cleavage face along the most exposed atoms on the top layer, representing the anisotropy of this facet. We found that the defined direction coincides with the longitudinal axis of the impurity layers, if there is one. For cleavage faces lacking such an axis, as in case A of Figure 2d, the impurity layers do not show obvious anisotropy. Thus the anisotropy of the crystal face does relate with that of the morphology of the impurity layers. This relationship holds for about 50 grains on the sample that have been scanned.

Now we turn to investigate the crystalline structure of graphene and the impurities. Electron diffraction pattern and dark field images have been proved an efficient beyond Raman spectroscopy[5] in the sense of resolution and identification of layer-number and stacking-sequence[3,6]. Diffraction pattern provides grain orientations[3]; corresponding dark field images provide mappings of grains and stacking-sequence[6,7,8]. Specifically, contrast in the second order dark field image is sufficient to identify layer-number variation[6].

To investigate the structure of our samples via TEM, we transferred them onto 10 nm thick silicon nitride membranes. Figure 3a is a bright-field TEM image of the sample showing an island similar to those seen in Fig. 1. Figure 3b shows the corresponding electron diffraction pattern. The diffraction pattern contains (1) bright main spots corresponding to monolayer graphene (2) a diffuse background from the amorphous $Si_3N_4$ membrane, and (3) sharp rings at the same radius as the monolayer graphene spots, with varying intensity along the rings, including some discrete spots. The pattern of intensity variation along the first-order and second-order rings has six-fold symmetry; the wave vector and symmetry of the sharp rings is entirely consistent with graphene indicating that the rings originate from crystalline graphitic carbon. Note that the possibility of the existence of additional carbon adatoms or/and nano-size isolated amorphous carbon cannot be excluded as these would be difficult to detect in diffraction. Dark field images of the first-order (Figure 3c) and second-order (Figure 3d) graphene diffraction spots are fairly uniform indicating that a monolayer of graphene corresponding to the main diffraction spots covers the whole image area. The second-order dark field image (Figure 3d) demonstrates that overwhelming majority of the additional material in the island region does not share the same orientation with the background monolayer graphene, though there are minority parts with a radial configuration and sizes ranging from nm to 100 nm that do align with the continuous monolayer. This observation is in contrast with bilayer graphene impurities in graphene grown on Cu[9], Ni[10,11] and Fe[12], and hardly Raman-detectable due to their small sizes.

Figure 3e shows an additional dark-field TEM image, with a larger aperture to include many diffraction spots on the sharp second-order ring. Here dark field TEM shows bright domains distributing throughout the whole surface inhomogeneously. The correspondence between the dark region in the bright field image Figure 3a and bright region in the dark field image Figure 3d and 3e confirms that the bright domains in Figure 3d and 3e mainly represent carbon impurities as an additional layer instead of crystalline defects in the continuous monolayer. This is the major finding of this paper: nanocrystalline graphene, misoriented with respect to the underlying continuous monolayer graphene, exists across the entire sample. The impurity graphene is inhomogeneously distributed, with a higher concentration in few-micron-sized island regions, but concentrations also detected outside the islands. This is consistent with what we observed by optical microscope and AFM.

We now turn to Raman spectroscopy of our Pt-grown graphene samples, using a 633 nm laser excitation source and a confocal microscope to map the Raman spectra at the submicron scale. Figure 4a (inset) shows a typical Raman spectrum of graphene grown on Pt similar to those examined in Figs. 1 and 3 and a reference sample grown on copper (with reduced carbon impurities as seen in TEM). It is immediately evident that the Raman spectrum for the Pt-grown graphene with carbon impurities is distinct from that reported for exfoliated graphene[5] or Cu-grown CVD graphene shown here or reported before[2]. The Pt-grown graphene exhibits a pronounced and extremely wide D peak and broad wings to the G peak. The spectrum could be consistent with the superposition of a graphene spectrum (with sharp G and 2D peaks) with a spectrum of nanocrystalline graphitic carbon (with very broad D and G peaks, and similar G and D intensities). The $I_D:I_G$ ratio attributed to the nanocrystalline graphene is about 1:1, corresponding to a grain size of order 10 nm[13]. The main panels of Figure 4a show the G and 2D peaks in greater detail. Excepting the broad background, the G peak of the impurity-decorated sample is similar in width, but slightly upshifted in position, compared to those of the Cu-grown sample (and typical exfoliated graphene samples). The impurity-decorated sample also shows a somewhat broader, upshifted, but still Lorentzian 2D peak.

Figure 4b shows Raman spectra of the impurity-decorated Pt-grown graphene taken along a line crossing an impurity island as shown in the inset. Raman spectra both on the island and off show the broad background in the D and G peak regions, indicating nanocrystalline graphitic carbon. The 2D peak remains Lorentzian everywhere, showing no evidence of large-sized Bernal-stacked graphite. An upshift of the 2D peak is associated with the island region, while the G peak is hardly changed in width or position. The consistency of the G peak excludes doping as the cause of the changes observed in the 2D peak[14]. Instead, the blueshift of 2D peak is in consistent with what was observed in turbostratic graphene[15,16] supporting our identification of the impurity carbon as graphitic but largely mis-aligned with the continuous monolayer graphene.

Figure 4c shows the electrical conductivity as a function of back-gate voltage of the impurity-decorated Pt-grown graphene transferred to a $SiO_2$/Si substrate. The field effect mobility is about 6000 $cm^2V^{-1}s^{-1}$, comparable to other graphene samples obtained by CVD[2, 17] showing much lower D peaks in Raman spectra. Again this is indicative that the D peak in our samples arises from defects in the carbon impurity layer, not in the continuous monolayer. We also explored the conductivity of a Hall bar region from which the impurity islands were excluded by etching and found only a slight mobility change (Figure 4c) indicating that the impurity island regions are not entirely responsible for the decreased mobility compared to exfoliated graphene on $SiO_2$[18,19].

**Discussion**

We have thoroughly characterized graphene grown by ambient-pressure CVD on Pt foils. We first find that many samples exhibit nano-crystalline graphitic carbon impurities in addition to a continuous single crystal monolayer graphene. The impurities are distributed inhomogeneously, especially concentrated in few-micron-sized island domains, but present everywhere on the sample.

It is important to understand whether the impurity carbon is deposited on top of the monolayer graphene or forms underneath. Our EBSD results show a strong correlation between the anisotropy of the impurity islands and the crystalline orientation of the Pt substrate (Figure 2), but not the crystalline orientation of the graphene which is continuous across many Pt grains. The correlation strongly suggests that the impurity islands form beneath the monolayer graphene at the Pt surface; given that the islands are not highly crystalline themselves the anisotropy probably results from anisotropic diffusion of carbon along the Pt. The detailed mechanism of the carbon cluster formation calls for further theoretical investigation[20,21]. Thus we can confirm the growth-from-below model for graphene on Pt[9].

Our investigation also points out difficulties and potential pitfalls in the standard characterization of multi-layered graphene by optical absorption and Raman spectroscopy. We find that amorphous carbon impurity islands can be easily mistaken for multi-layer regions in optical images. Characterization by TEM is thus important for reports of new synthesis methods for graphene, since it gives the most accurate crystalline information. The Raman spectra of our sample shows a very large and broad D feature normally associated with highly defected carbon[13]. Here we have associated the enhancement of D peak with carbon impurities; the high electronic quality of our sample excludes the possibility that the defects responsible for the D band are in the continuous monolayer graphene. The effects of carbon impurities on the electronic transport is a topic for further investigation.

**Methods**

The graphene samples studied here were grown at 1000 °C on a 0.2 mm thick platinum foil under flowing ambient pressure gas mixture of 700 sccm hydrogen and 5 sccm methane[1].

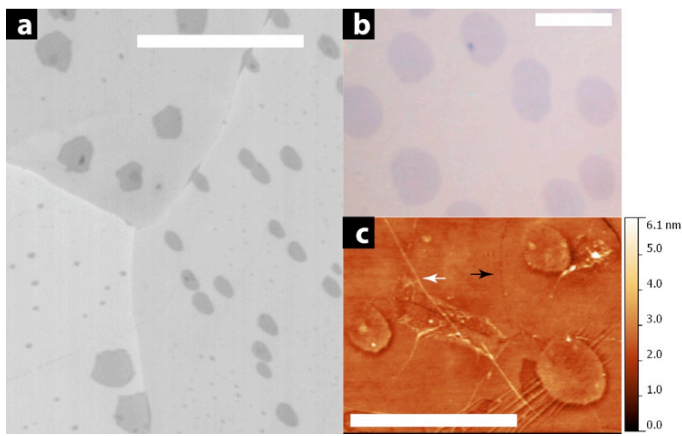

**Figure 1.** Scanning electron micrograph(a), optical micrograph(b), and atomic force micrograph (c) of graphene grown by ambient-pressure chemical vapor deposition on Pt foil. Image (a) is as-grown sample on Pt foil; the scale bar is 5 µm. The three large areas of different contrast are three crystalline grains of the Pt foil. Image (b) is taken after transfer of graphene to $SiO_2$/Si substrate; the scale bar is 10 µm. Image (c) is taken on graphene on poly methylmethacrylate transfer layer transferred "upside-down" to $SiO_2$/Si. The scale bar is 5 µm. Raised and lowered wrinkles are indicated by white and black arrows respectively. In all cases few-micron-sized islands of different contrast are evident.

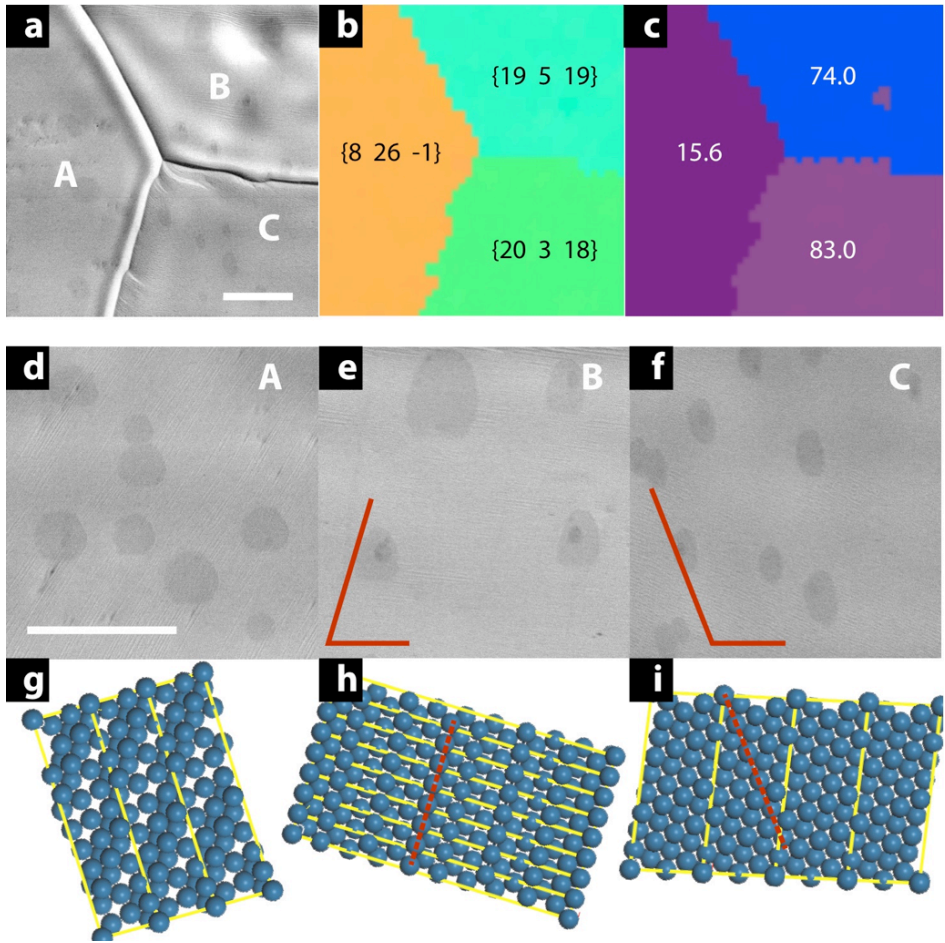

**Figure 2.** The SEM image (a) shows platinum substrate with 3 grains A, B, and C of different orientations, whose IPF map and Euler angles map are shown in (b) and (c). The plane direction {h k l} and the last Euler angle $\phi_2$ are shown in the corresponding map. The atomic arrangement are shown respectively in (g), (h), and (i) below detailed SEM images in (d), (e), and (f) marked with directions, if exist, defined in the text. The angles formed by the axes and the RD are also shown in the SEM images to guide eyes. The scale bars in (a) and (b) are 5 μm.

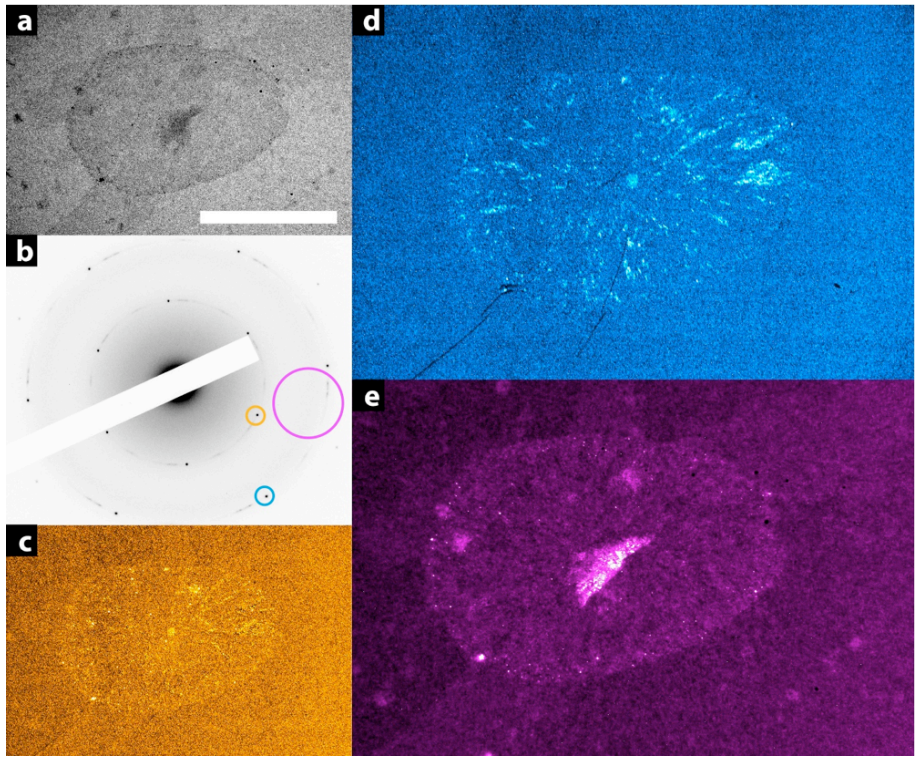

**Figure 3.** Transmission electron microscopy and diffraction of Pt-grown graphene. (a) Bright field image with scale bar of 1 μm. (b) Electron diffraction pattern corresponding to sample in (a). (c-e) Dark field images corresponding to orange (c), blue (d) and magenta (e) apertures indicated in (b).

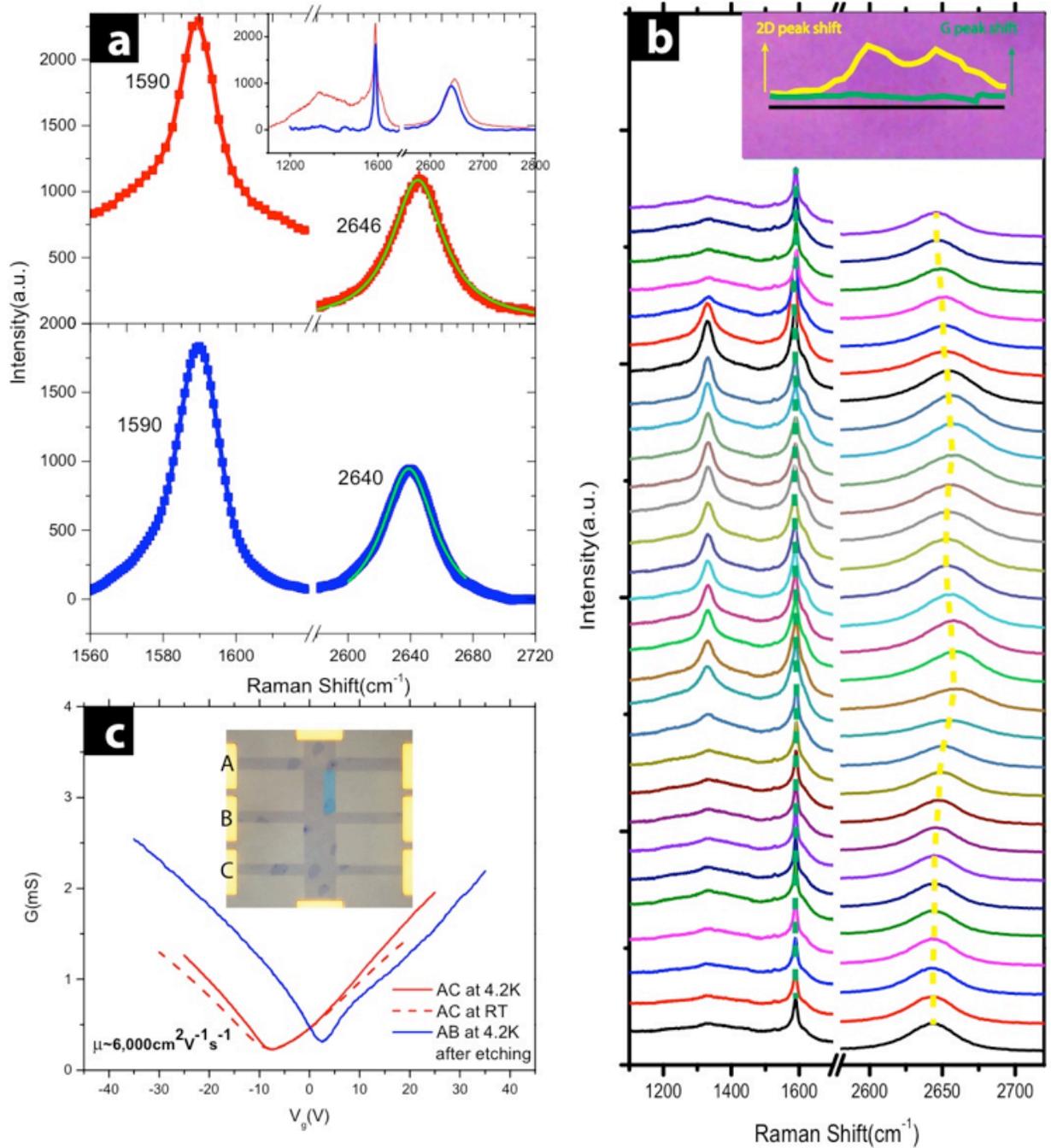

**Figure 4.** Raman spectroscopy and electrical characterization of Pt-grown graphene. (a) Raman spectra of a Pt-grown graphene sample with impurities similar to Figures 1 and 2 is shown in red, and for comparison Raman spectra from a Cu-grown graphene sample without impurity regions is shown in blue. Inset shows full spectra while main panel shows the G and 2D peak regions, with peak positions in wavenumbers (cm$^{-1}$). Lorentzian fits to the 2D peaks are shown in green. (b) Raman spectra taken at points with spacing of 0.31 μm along the 9 μm long black segment in the optical image in inset. The position or blueshift of the 2D peak is extracted from Raman spectral and projected onto the real-position optical image. The length of the yellow and green arrows is 15.32 cm$^{-1}$. Note that the maximum blueshift happens inside the edge of the island due to μm resolution of Raman microscopy. (c) Conductivity as a function of back gate voltage for a Pt-grown graphene sample with impurities on 300 nm $SiO_2$/Si substrate. An image of the Hall bar device is shown in the inset; the top and bottom electrodes are used to source and sink current. The width of the Hall bar is 10 μm. The red data curves correspond to voltages measured between electrodes A and C at 4.2 K and room temperature (RT) before etching. The blue curve corresponds to voltage probes A and B after the blue-shadowed region in the inset optical image was removed by etching.

**Acknowledgements**

We acknowledge support from the ONR-MURI program, and M.S.F. acknowledges support from an Australian Research Council Laureate Fellowship. We also acknowledge the support of the Maryland NanoCenter and use of Shared Experimental Facilities of the University of Maryland MRSEC (NSF grant DMR 05-20741) including the NispLab and of the facilities and the assistance of Xi-Ya Fang of at the Monash Centre for Electron Microscopy.

**Author Contributions**

Jinglei Ping and Michael Sears Fuhrer conceived the experiments, discussed the results and co-wrote the paper. Jinglei Ping performed the experiments and analyzed the data.

**Additional Information**

The authors declare no competing financial interests.